\documentclass[conference]{IEEEtran}
\IEEEoverridecommandlockouts

\usepackage{amsthm}
\usepackage{thm-restate}

\usepackage{cite}
\usepackage{amsmath,amssymb,amsfonts}

\usepackage{algorithm}
\usepackage{algorithmic}

\usepackage{graphicx}
\usepackage{textcomp}
\usepackage{xcolor}

\usepackage{subcaption}

\usepackage{enumerate}
\usepackage{booktabs}

\newcommand{\real}{\mathbb{R}}

\newcommand{\cancel}[1]{}
\newcommand{\supp}{\mathrm{supp}}
\newcommand{\norm}[1]{\|#1\|}
\newcommand{\ex}{\mathrm{E}}
\newcommand{\var}{\mathrm{Var}}
\newcommand{\cov}{\mathrm{Cov}}

\newtheorem{theorem}{Theorem}
\newtheorem{lemma}[theorem]{Lemma}

\def\BibTeX{{\rm B\kern-.05em{\sc i\kern-.025em b}\kern-.08em
    T\kern-.1667em\lower.7ex\hbox{E}\kern-.125emX}}
\begin{document}

\title{Sparse Signal Recovery from Random Measurements\thanks{Research supported by Research Grants Council, Hong Kong, China (project no.~16203718).}
}

\author{\IEEEauthorblockN{1\textsuperscript{st}  Man Ting Wong}
\IEEEauthorblockA{
\textit{Hong Kong University of Science and Technology}\\
Hong Kong, China \\
mtwongaf@connect.ust.hk}
\and
\IEEEauthorblockN{2\textsuperscript{nd}Siu-Wing Cheng}
\IEEEauthorblockA{
\textit{Hong Kong University of Science and Technology}\\
Hong Kong, China \\
scheng@cse.ust.hk}

}

\maketitle

\begin{abstract}
Given the compressed sensing measurements of an unknown vector $z \in \real^n$ using random matrices, we present a simple method to determine $z$ without solving any optimization problem or linear system.
Our method uses $\Theta(\log n)$ random sensing matrices in $\real^{k \times n}$ and runs in $O(kn\log n)$ time, where $k = \Theta(s\log n)$ and $s$ is the number of nonzero coordinates in $z$. We adapt our method to determine the support set of $z$ and experimentally compare with some optimization-based methods on binary signals.  
\end{abstract}


\section{Introduction}

Given a signal $z \in \real^n$ with $s \ll n$ nonzero coordinates, it has been shown that one can recover $z$ with $k \ll n$ measurements under some conditions.  Such recovery methods are known as \emph{compressed sensing}~\cite{bar08,rupe,cs} with numerous applications. The value of $s$ is not more than 8\% of $n$ in many cases~\cite{ssec,spiv}.  It has also been argued~\cite{cs} that $O(\sqrt{n}\log^3 n)$ measurements suffice to reconstruct certain images with $n$ pixels, that is, $s = o(n)$. 

In this paper, we adopt the \emph{linear measurement model} in which $k$ measurements are bundled as a $k$-dimensional vector $b = Az$, where $A$ is a matrix in $\real^{k \times n}$.  We are not concerned with the actual sensing mechanism that produces $b$ according to the linear measurement model.  We are only concerned with the recovery of $z$ given $A$ and $b$.

Given $A$ and $b$, it has been proved that $z$ can recovered exactly by minimizing $\frac{1}{2}\norm{Ax-b}^2 + \eta\norm{x}_1$ over all $x \in \real^n$ with an appropriate choice of $\eta > 0$, provided that $A$ satisfies the \emph{restricted isometric property \emph{(RIP)} of order $3s$}~\cite{bar08,rupe,cs}.  The matrix $A$ has order $3s$ if there exists $\delta \in (0,1)$ such that $(1-\delta)\norm{x}^2 \leq \norm{Ax}^2 \leq (1+\delta)\norm{x}^2$ for all vectors $x \in \real^n$ with at most $3s$ nonzero coordinates.  A popular way to construct a matrix $A \in \real^{k \times n}$ is to draw each entry of $A$ independently from $\mathcal{N}(0,1/k)$, i.e., each matrix entry is a Gaussian variable with mean zero and variance $1/k$. Such a matrix satisfies the RIP of order $s$ with probability at least $1 - 2e^{-\Theta(s)}$, provided that $k \geq Cs\ln(n/s)$ for some appropriate constant $C$~\cite{bar08}.  Random matrices are resilient to noise as well.  On the other hand, it has been observed in the literature that it can be resource-demanding to solve $\min\frac{1}{2}\norm{Ax-b}^2 + \eta\norm{x}_1$ or other similar optimization problems for large $n$.  Also, there is no efficient algorithm to check whether a random matrix satisfies the RIP.

\subsection{Our results}

We examine the following question: given the random matrices and measurements, can we determine $z$ \emph{without} solving any optimization problem or linear system?
We present such an algorithm. The most involved step is computing the median of a set of numbers.  

We use $O(\log n)$ random matrices in $\real^{k \times n}$, where $k = \Omega(s\log n)$.  Every matrix entry is an iid Gaussian sample of mean zero and variance $1/k$.  We model noise by adding to each measurement an iid Gaussian sample of mean zero and variance $\sigma_w^2$.   The recovery algorithm runs in $O(kn\log n)$ time.  It holds with probability $1-O(1/n)$ that every recovered coordinate has an additive error less than $2\sigma_w$ for a sufficiently large $n$.  We can also ensure that coordinates not in the support set are zero.

We adapt our method to determine the support set of $z$ and run experiments on binary signals together with several optimization-based methods, including gradient projection for sparse reconstruction (GPSR)~\cite{gpsr}, dynamic working set method (DWS)~\cite{DWS}, orthogonal matching pursuit (OMP)~\cite{OMP}, binary iterative hard thresholding (BIHT)~\cite{JLBB2013b}, and normalized BIHT (NBIHT)~\cite{FJPY2022}. These optimization-based methods can work with arbitrary sensing matrices, although the convergence of BIHT and NBIHT is analyzed using properties that can be guaranteed by random matrices.  

Our accuracy is higher than GPSR and DWS and comparable with OMP.  BIHT and NBIHT are designed for the harder 1-bit compressed sensing problem, which can also be used to determine the support set of $z$.  So we also compare our method with BIHT and NBIHT.  As expected, our accuracy is much higer because the 1-bit compressed sensing problem is harder.  When $z$ is binary, our method achieves a speedup of $1.95\times$ to $23.92\times$ (or higher) relative to GPSR, DWS, and OMP.

\subsection{Related works}

There has been a lot of work on constructing the sensing matrix $A$ deterministically that are based on codes, designs, algebraic curves, geometry over finite fields, and additive combinatorics (e.g.~\cite{AMM2012,BDFKK2011,BCHO2017,LGGZ2012,LG2014}).  These matrices are usually sparse which leads to improved efficiency and facilitates the use of compressed sensing for larger problems.  Nevertheless, such methods still resort to solving an optimization problem to recover the unknown signal.

The minimization of $\frac{1}{2}\norm{Ax - b}^2 + \eta\norm{x}_1$ is generally known as Lasso~\cite{lasso}.  Several solvers have been proposed, including GPSR~\cite{gpsr}, IST~\cite{dd04}, L1\_LS~\cite{kse07}, L1-magic~\cite{l1magic}, and the homotopy method~\cite{ddy08}.  In compressed sensing experiments, GPSR outperforms IST and L1\_LS~\cite{gpsr}, and L1\_LS runs faster than L1-magic and the homotopy method~\cite{l1magic}.  There are also coordinate descent algorithms such as glmnet~\cite{glmlassopaper} and scikit-learn~\cite{scikit} for solving large convex optimization problems with sparse solutions.  There are also fast working set algorithms, including Picasso~\cite{picasso}, Blitz~\cite{Blitz}, Fireworks~\cite{fireworks}, Skglm~\cite{skglm}, Celer~\cite{celer2}, and DWS~\cite{DWS}.  In Lasso experiments, Blitz runs faster than L1\_LS and glmnet~\cite{Blitz}, Celer performs better than Blitz and scikit-learn~\cite{celer2}, and Skglm runs faster than Celer, Blitz, and Picasso~\cite{skglm}.  When $s$ is less than 8\% of $n$, DWS outperforms GPSR, Skglm, and Celer~\cite{DWS}.

There are dedicated algorithms for signal recovery.  The popular ones include OMP~\cite{OMP} and approximate message passing (AMP)~\cite{AMP1,AMP2}.  AMP expects the input signals to follow a distribution which is given to the algorithm.  When such a distribution is absent (e.g.~the input signals are arbitrary), the performance is adversely affected.  There are some iterative methods to improve the recovery efficiency for binary signals in ~\cite{NR2012,SLVYZ2015} and for general signals in~\cite{L2016}.  For general signals, an optimization problem may need to be solved in~\cite{L2016} in some intermediate steps.
In~\cite{NR2012,SLVYZ2015}, a measurement involves a small constant $\beta$ of nonzero coordinates of $z$.  This is important because a table of size $2^\beta$ is created, which is feasible only if $\beta$ is small.

There is the related problem of 1-bit compressed sensing.  In this setting, only the signs of the coordinates of $Az$ are given, and the problem is to recover an approximation of $z$ or just the support set of $z$.  Binary iterative hard thresholding (BIHT)~\cite{JLBB2013b} is the widely celebrated algorithm for 1-bit compressed sensing.   A closely related algorithm is the normalized BIHT (NBIHT)~\cite{FJPY2022}.  Various theoretical progresses have been obtained, including convergence of BIHT and NBIHT~\cite{MM2024,FJPY2022}, bounds for the sample size for 1-bit compressed sensing~\cite{ABK2017,MMP2023}, and computing the support set of $z$ or an approximation of $z$ in $o(n)$ time~\cite{YZG2025}.

\subsection{Notations.}  

Given a vector $x$, we use $x_i$ to denote its $i$-th coordinate.  Let $z$ denote an unknown signal in $\real^n$.  Let $\supp(z) = \{ i \in [n] : z_i \not= 0\}$.  The \emph{sparsity} of $z$ is $s = |\supp(z)|$.  One way to form a random matrix in $\real^{k \times n}$ is to draw each matrix entry independently from $\mathcal{N}(0,1/k)$.  When we say that a matrix belongs to $\mathcal{N}^{k \times n}(0,1/k)$, it means that the matrix is generated as described above.  When we say that a vector belongs to $\mathcal{N}^k(0,\lambda)$, it means that it is a $k$-dimensional vector in which each coordinate is drawn independently from $\mathcal{N}(0,\lambda)$.  

Let $r_0$ be a parameter that will be specified later.
We assume that the input provides a set of random matrices $A^{(r)} \in \mathcal{N}^{k \times m}(0,1/k)$ for $r \in [r_0]$.  Let $z \in \real^n$ be an unknown signal. For each $A^{(r)}$, the input provides $k$ measurements of~$z$, bundled in a vector $b^{(r)} = A^{(r)}z + w$ for some noise $w \in \mathcal{N}^k(0,\sigma_w^2)$. Neither the sparsity $s$ nor $\sigma_w$ is known to us.

\section{Algorithms and Analysis}
\label{sc:algorithm}

We present our algorithms and state the key results in the analysis.  The missing proofs are deferred to the appendix.  We use $\ex[\cdot]$, $\var[\cdot]$, and $\cov[\cdot,\cdot]$ to denote the expected value, variance, and covariance of random variables, respectively.

Algorithm~\ref{alg:recovery} describes the basics of our recovery method.

\begin{algorithm}
\caption{Basic Recovery Algorithm}
\label{alg:recovery}
\begin{algorithmic}[1]
\FOR{$r = 1$ to $r_0$}
    \STATE $v^{(r)} \gets (A^{(r)})^{\top} b^{(r)}$
\ENDFOR
\STATE initialize a zero vector $\hat{z}$
\FOR{$i = 1$ to $n$}
    \STATE $\hat{z}_i \gets \mathrm{median}\bigl\{v_i^{(r)} : r \in [r_0]\bigr\}$
\ENDFOR
\STATE \textbf{return} $\hat{z}$
\end{algorithmic}
\end{algorithm}

Lemma~\ref{lem:tech1} below shows that the coordinates of $v^{(r)}$ are independent, good estimators of the coordinates of $z$.  It is proved by direct calculations.

\begin{lemma}
\label{lem:tech1}
We write $v_i^{(r)}$ as $v_i$ for convenience.

\begin{enumerate}[{\em (i)}]

\item $\forall\,\, i \in [n]$, $\ex[v_i] = z_i$.

\item $\forall\,\, i,j \in [n]$, if $i \not= j$, then $\cov(v_i,v_j) = 0$.

\item $\forall\,\, i \in [n]$, $\var[v_i] = \frac{1}{k}z_i^2 + \frac{1}{k}\norm{z}^2 + \sigma_w^2$.

\end{enumerate}
\end{lemma}

The guarantees of Algorithm~\ref{alg:recovery} can be obtained using Lemma~\ref{lem:tech1} and Chebyshev's inequality.

\begin{theorem}
\label{thm:recovery-1}
Assume that $k = \Omega(s\log n)$ and the upper bound of $\max_{i \in \supp(z)} |z_i|$ is fixed independent of $n$.  Algorithm~\ref{alg:recovery} returns $\hat{z} \in \real^n$ in $O(knr_0)$ time such that when $n$ is sufficiently large, $|\hat{z}_i - z_i| < 2\sigma_w$ for all $i \in [n]$ with probability at least $1-ne^{-r_0/2}$.
\end{theorem}
\begin{proof}
The running time is clearly $O(knr_0)$.  Let $\theta_i^2 = \var\bigl[v_i^{(r)}\bigr]$.  We have $\Pr\bigl[|v_i^{(r)} - z_i| \geq \sqrt{e}\theta_i\bigr] \leq 1/e$ by Chebyshev's inequality.  So the median $v_i^{(r)}$ has an error $\sqrt{e}\theta_i$ or more with probability at most $e^{-r_0/2}$.  Since $k = \Omega(s\log  n)$, $\norm{z}^2/k \leq O(\max_{i\in \supp(z)}|z_i|^2/\log n)$ which is less than $0.2\sigma_w^2$ for a large enough $n$.  So is $|z_i|^2/k$.  Therefore, $\sqrt{e}\theta_i \leq (\sqrt{1.4e}) \sigma_w < 2\sigma_w$.
\end{proof}

To make $\hat{z}_i = 0$ for all $i \not\in \supp(z)$, we must estimate $\norm{z}^2 + k\sigma_w^2$ by Lemma~\ref{lem:tech1}(iii).  Lemma~\ref{lem:tech2} below shows how to do it.  The expectation and variance of $\norm{b^{(r)}}^2$ are obtained by direct calculations.  The probabilistic range of $\sigma^2$ is proved using Chebyshev's inequality and Chernoff bound.

\begin{lemma}
\label{lem:tech2}
We write $b^{(r)}$ as $b$ for convenience.
\begin{enumerate}[{\em (i)}]

\item $\ex\bigl[\norm{b}^2\bigr] = \norm{z}^2 + k\sigma_w^2$.

\item $\var\bigl[\norm{b}^2\bigr] = \frac{2}{k}\left(\norm{z}^2 + k\sigma_w^2\right)^2$.

\end{enumerate}
Let $\sigma^2 = \mathrm{median}\{\norm{b^{(r)}}^2 : r \in [r_0]\}$.  It holds with probability more than $1-2e^{-r_0/72}$ that $\sigma^2$ lies strictly between $(1 -\sqrt{6/k}\bigr)\cdot\left(\|z\|^2 + k\sigma_w^2\right)$ and $\bigl(1 + \sqrt{6/k}\bigr) \cdot \left(\|z\|^2 + k\sigma_w^2\right)$.
\end{lemma}

Lemma~\ref{lem:class1} below shows that $2\sigma/\sqrt{k}$ is a good threshold for deciding whether $i \in \supp(z)$ under the assumption that $\min_{i \in \supp(z)} |z_i| \geq 6\sigma/\sqrt{k}$.  It is also proved using Chebyshev's inequality and Chernoff bound.

\begin{lemma}
\label{lem:class1}
Assume that $k \geq 54$, $\sigma > (1-\sqrt{6/k})^{1/2}\cdot \sqrt{\norm{z}^2 + k\sigma_w^2}$, and $\min_{i \in \supp(z)} |z_i| \geq 6\sigma/\sqrt{k}$.  It holds with probability at least $1-2ne^{-r_0/540}$ that for $i \in [n]$, if $i \not\in \supp(z)$, then $\hat{z}_i < 2\sigma/\sqrt{k}$; otherwise, $\hat{z}_i > 2\sigma/\sqrt{k}$.
\end{lemma}

We use Lemma~\ref{lem:class1} to suppress the coordinates that are not in $\supp(z)$.  We need to use another $r_0$ random matrices and measurements to compute $\sigma^2$ by Lemma~\ref{lem:tech2}.

\begin{algorithm}
\caption{Recovery Algorithm}
\label{alg:recovery-2}
\begin{algorithmic}[1]
\STATE run Algorithm 1 to obtain $\hat{z}$
\STATE $\sigma^2 \gets \text{median}\left\{ \lVert b^{(r)} \rVert^2 : r \in [r_0+1,2r_0] \right\}$.
\FOR{$i = 1$ to $n$}
  \IF{$\hat{z}_i < 2\sigma/\sqrt{k}$}
      \STATE $\hat{z}_i \gets 0$
  \ENDIF
\ENDFOR
\end{algorithmic}
\end{algorithm}

\begin{theorem}
Assume that $k = \Omega(s \log n)$, $\min_{i \in \supp(z)} |z_i| \geq 15\sigma_w$, and the upper bound of $max_{i \in \supp(z)} |z_i|$ is fixed independent of $n$.  Let $r_0 = 1080\ln n$. Algorithm~\ref{alg:recovery-2} returns $\hat{z} \in \real^n$ in $O(kn\log n)$ time, and it holds with probability $1-O(1/n)$ that when $n$ is sufficiently large, for all $i \in [n]$, if $i \in \supp(z)$, then $|\hat{z}_i-z_i| < 2\sigma_w$; otherwise, $\hat{z}_i = 0$.
\end{theorem}
\begin{proof}
The running time follows from the description of the algorithm.  Since $r_0 = 1080\ln n$, the probability bounds in Theorem~\ref{thm:recovery-1} and Lemmas~\ref{lem:tech2} and~\ref{lem:class1} are $1-O(1/n)$.  

By Lemma~\ref{lem:tech2}, $\sigma > (1-\sqrt{6/k})^{1/2}\cdot \sqrt{\norm{z}^2 + k\sigma_w^2}$ and $\sigma^2/k < \bigl(1+\sqrt{6/k}\bigr)(\frac{1}{k}\norm{z}^2 + \sigma_w^2)$.  We have seen in the proof of Theorem~\ref{thm:recovery-1} that $\frac{1}{k}\norm{z}^2 + \sigma_w^2 \leq 1.2\sigma_w^2$. Therefore, $\min_{i \in \supp(z)} |z_i| > 6\sigma/\sqrt{k}$ for large enough $n$ because $\min_{i \in \supp(z)} |z_i| \geq 15\sigma_w$ by assumption.  It follows that Lemmas~\ref{lem:tech2} and~\ref{lem:class1} are applicable.  

So  Algorithm~\ref{alg:recovery-2} sets $\hat{z}_i = 0$ for all $i \not\in \supp(z)$ with probability $1-O(1/n)$.  It follows from Theorem~\ref{thm:recovery-1} that $|\hat{z}_i - z_i| < 2\sigma_w$ for all $i \in \supp(z)$.
\end{proof}

If we are only interested in determining $\supp(z)$, we can replace the repeated median computations in Algorithm~\ref{alg:recovery} by simple counting.  Refer to Algorithm~\ref{alg:support-estimation}.  Since we only to need to compare with $2\sigma/\sqrt{k}$ to classify $i \in [n]$, we can give a concrete lower bound for $k$ that is independent of $n$.

\begin{algorithm}
\caption{Support Set Determination}
\label{alg:support-estimation}
\begin{algorithmic}[1]
\FOR{$r = 1$ to $r_0$}
    \STATE $v^{(r)} \gets (A^{(r)})^{\top} b^{(r)}$
\ENDFOR
\STATE $\sigma^2 \gets \operatorname{median}\left\{ \lVert b^{(r)} \rVert^2 : r \in [r_0+1,2r_0] \right\}$
\STATE initialize $S \gets \emptyset$.
\FOR{$i = 1$ to $n$}
    \STATE $c_i \gets \bigl| \bigl\{ r \in [r_0] : |v_i^{(r)}| \geq 2\sigma/\sqrt{k} \bigr\} \bigr|$
    \IF{$c_i \geq \lceil r_0/2 \rceil$}
        \STATE $S \gets S \cup \{i\}$.
    \ENDIF
\ENDFOR

\STATE \textbf{return} $S$ as an estimate of $\supp(z)$.
\end{algorithmic}
\end{algorithm}

\begin{theorem}
Let $\Delta = \max_{i\in \supp(z)} |z_i| / \min_{i \in \supp(z)} |z_i|$.  Let $r_0 = 1080\ln n$.  Suppose that $\min_{i \in \supp(z)} |z_i| \geq 15\sigma_w$ and $k \geq 225s\Delta^2$.  Algorithm~\ref{alg:support-estimation} runs in $O(kn\log n)$ time, and the output is $\supp(z)$ with probability $1-O(1/n)$. 
\end{theorem}
\begin{proof}
We first show that $\min_{i \in \supp(z)} |z_i| \geq 6\sigma/\sqrt{k}$.  Assume that $\min_{i \in \supp(z)} |z_i| = 1$.  So $\Delta = \max_{i \in \supp(z)} |z_i|$.  As $k \geq 225s\Delta^2$, we have $(6/\sqrt{k})\bigl(1 + 6/\sqrt{k}\bigr)^{1/2} \cdot \norm{z} \leq (6/\sqrt{k})\bigl(1+6/\sqrt{k}\bigr)^{1/2} \cdot \sqrt{s}\Delta < 0.5$.  As $15\sigma_w \leq \min_{i\in\supp(z)} |z_i| = 1$, we have $(6/\sqrt{k})\bigl(1 + 6/\sqrt{k}\bigr)^{1/2} \cdot \sqrt{k}\sigma_w < 0.5$.  By Lemma~\ref{lem:tech2}, $\sigma < (1 + \sqrt{6/k})^{1/2} \cdot \bigl(\norm{z}^2 + k\sigma_w^2\bigr)^{1/2} \leq (1+6/\sqrt{k})^{1/2}\cdot \norm{z} + (1+6/\sqrt{k})^{1/2}\cdot \sqrt{k}\sigma_w$ with probability at least $1 - O(1/n)$.  Thus, $6\sigma/\sqrt{k} < (6/\sqrt{k})(1+6/\sqrt{k})^{1/2}\cdot\norm{z} + (6/\sqrt{k})(1+6/\sqrt{k})^{1/2}\cdot\sqrt{k}\sigma_w < 1 = \min_{i\in\supp(z)} |z_i|$.  As a result, Lemmas~\ref{lem:tech2} and~\ref{lem:class1} are applicable.   Since $r_0 = 1080\ln n$, the probability bounds in these two lemmas are $1-O(1/n)$.  It follows that Algorithm~\ref{alg:support-estimation} correctly identifies $\supp(z)$ with probability $1-O(1/n)$.
\end{proof}

\section{Experimental results}

\subsection{Setup}

We implemented Algorithm~\ref{alg:support-estimation} and run experiments on synthetic binary signals.  We compare its effectiveness in determining the support set with several optimization-based methods.  All experiments are conducted on a machine equipped with a 12th Gen Intel i9-12900KF CPU (3.19\,GHz), 64\,GB of RAM, and MATLAB R2023b. 

We consider three signal sizes, $n \in \{2000, 4000, 8000\}$, with sparsity $s$ ranging from $1\%$ to $8\%$ of $n$ as in~\cite{ssec,spiv}. We generate $z \in \{0,1\}^n$ by choosing $s$ coordinates  uniformly at random to be 1, and the other coordinates are zeros. 

The measurement matrices are independently sampled from $\mathcal{N}^{k \times n}(0, 1/k)$, and measurements are taken as $b = Az + w$, where the noise vector $w \sim \mathcal{N}^k(0, \sigma_w^2/k)$ with $\sigma_w = 0.1$.   The number of measurements is set to be $k = \lceil 2s \ln n \rceil$ as in~\cite{srfr}. With this setting of $k$, $z$ can be recovered by minimizing $\frac{1}{2}\norm{Az-b}^2 + \eta\norm{z}_1$ for $A\in \mathbb{R}^{k\times n}$ with $\eta=0.1$ as in~\cite{gpsr}.

For each combination $(s,n)$, we perform 273 independent trials, i.e., generate 273 random binary signals, take their measurements, and determine their support sets from the measurements using our algorithm and several other algorithms:
\begin{itemize}
    \item \textbf{RAND}: The implementation of Algorithm~\ref{alg:support-estimation} with $r_0 = \lceil \ln n\rceil$.  The measurements are taken using $2r_0$ random matrices.

    \item \textbf{GPSR}: a widely used solver for $\ell_1$-minimization problems~\cite{gpsr}. In compressed sensing experiments, it outperforms several other methods, including IST~\cite{dd04} and L1\_LS~\cite{kse07}, L1-magic~\cite{l1magic}, and the homotopy method~\cite{ddy08}. 
    
    \item \textbf{DWS}: a dynamic working set method developed by the authors~\cite{DWS}.  When $s$ is less than 8\% of $n$, DWS outperforms the fast working set algorithms Skglm~\cite{skglm} and Celer~\cite{celer}, which in turn give better performances than Picasso~\cite{picasso}, Blitz~\cite{Blitz}, and Fireworks~\cite{fireworks}.
    
    \item \textbf{OMP}: a popular greedy algorithm for compressed sensing and signal recovery~\cite{OMP}.

    \item \textbf{BIHT}: a widely celebrated algorithm for 1-bit compressed sensing~\cite{JLBB2013b}.

    \item \textbf{NBIHT}: the normalized BIHT in which every intermediate solution is normalized before computing the next intermediate solution~\cite{FJPY2022}. It may help BIHT to converge faster in later iterations.
    
\end{itemize}

GPSR, DWS, OMP, BIHT, and NBIHT use only one sensing matrix and one measurement vector, whereas RAND processes $2r_0$ sensing matrices and $2r_0$ measurement vectors.  

An upper bound on sparsity is required by OMP, BIHT, and NBIHT, so we pass the value of $s$ to them.  

BIHT and NBIHT are designed for one-bit compressed sensing when only the signs of $Az$ are given.  If a coordinate of $Az$ is postive, we convert it to $+1$; otherwise, we convert it to $-1$.  One-bit compressed sensing may be a harder problem, but we include the comparisons with BIHT and NBIHT for completeness, as they can also determine the support set of $z$.  

Although we generate our binary signals randomly in our experiments, this is done out of convenience.  All algorithms above are designed to handle arbitrary signals and do not exploit this prior distribution.

\subsection{Accuracy}

Let \(x\) be the solution returned by a solver, and let \(z\) be the ground truth of the unknown signal. To demonstrate the performance of support prediction for the algorithms, we define the ratio $R=|\supp(x)\cap\supp(z)|/|\supp(x)\cup\supp(z)|$
as the accuracy. This ratio equals \(1\) when the predicted support is identical to the ground truth. It decreases when \(\supp(x)\) does not contain all indices in \(\supp(z)\) or when it includes false predictions outside \(\supp(z)\). 
Across all methods, $R$ has an expected value at least 0.5 and a variance below \(0.01\) over the trials.  So $R$ is an effective accuracy indicator.

\begin{figure}[htbp]
    \centering
    \begin{minipage}[b]{0.5\textwidth}
        \centering
        \includegraphics[width=0.9\linewidth]{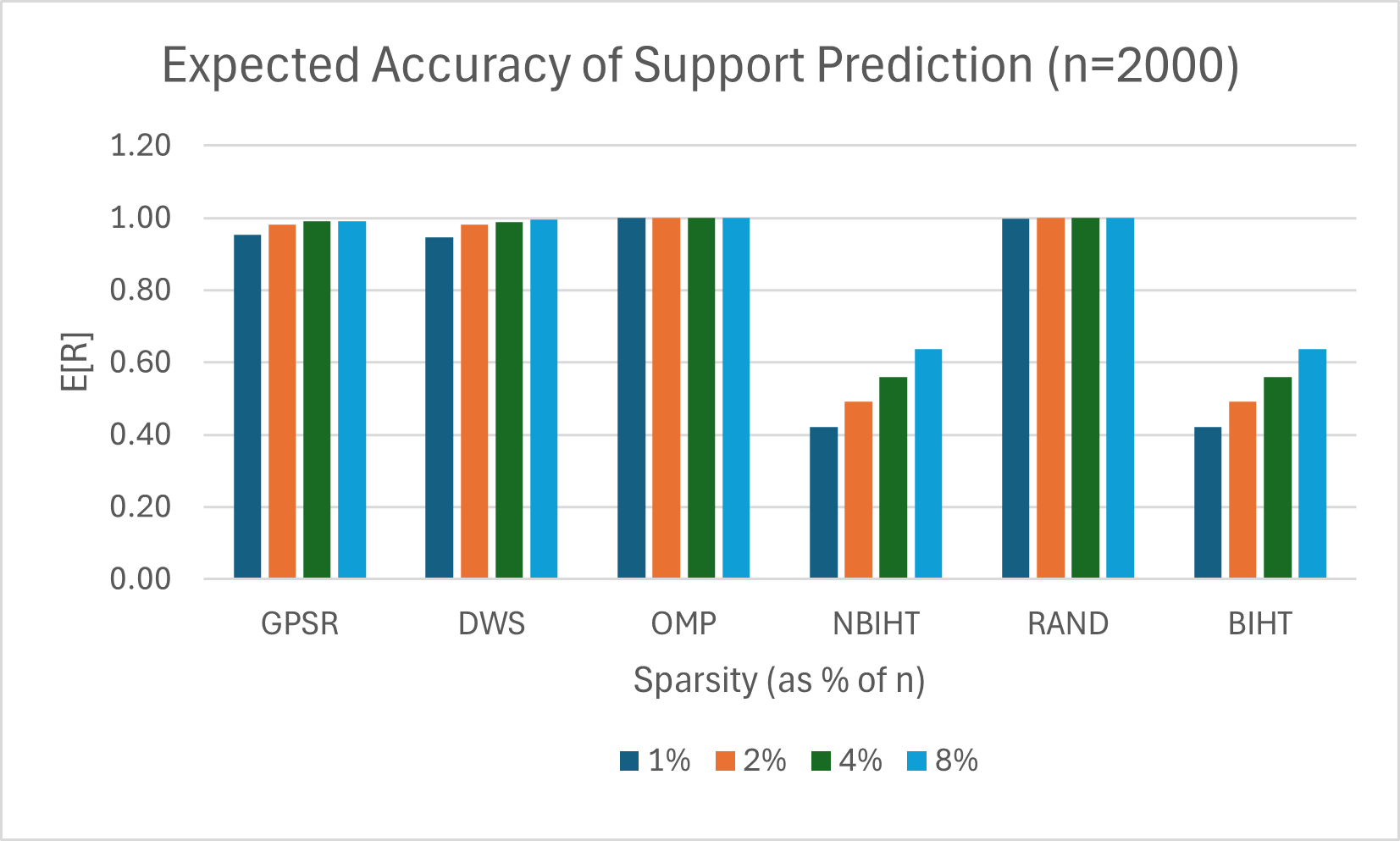}
        \subcaption{Performance for $n=2000$.}\label{fig:sp1}
    \end{minipage}
    \hfill
    \begin{minipage}[b]{0.5\textwidth}
        \centering
        \includegraphics[width=0.9\linewidth]{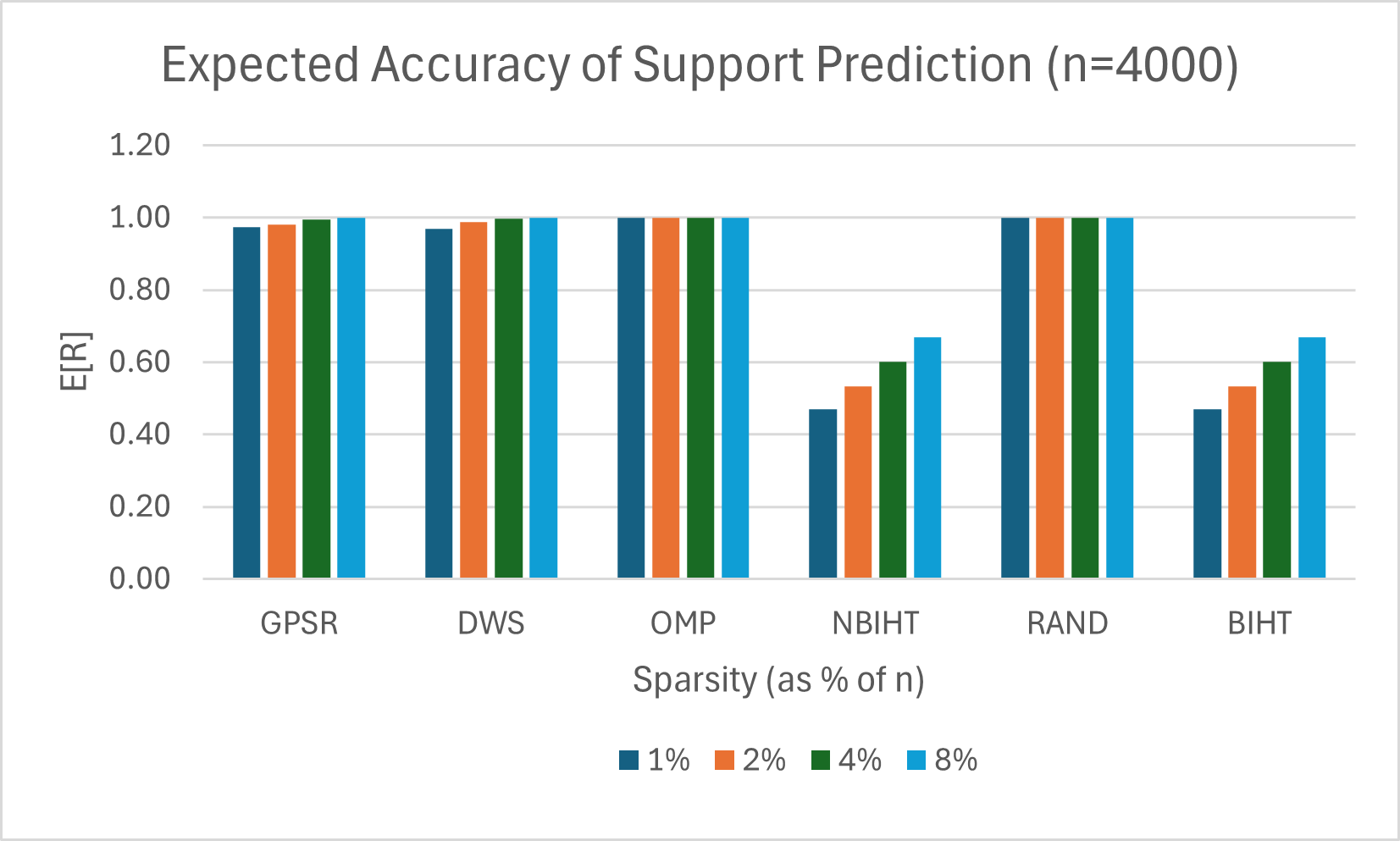}
        \subcaption{Performance for $n=4000$.}\label{fig:sp2}
    \end{minipage}
    \hfill
   \begin{minipage}[b]{0.5\textwidth}
        \centering
        \includegraphics[width=0.9\linewidth]{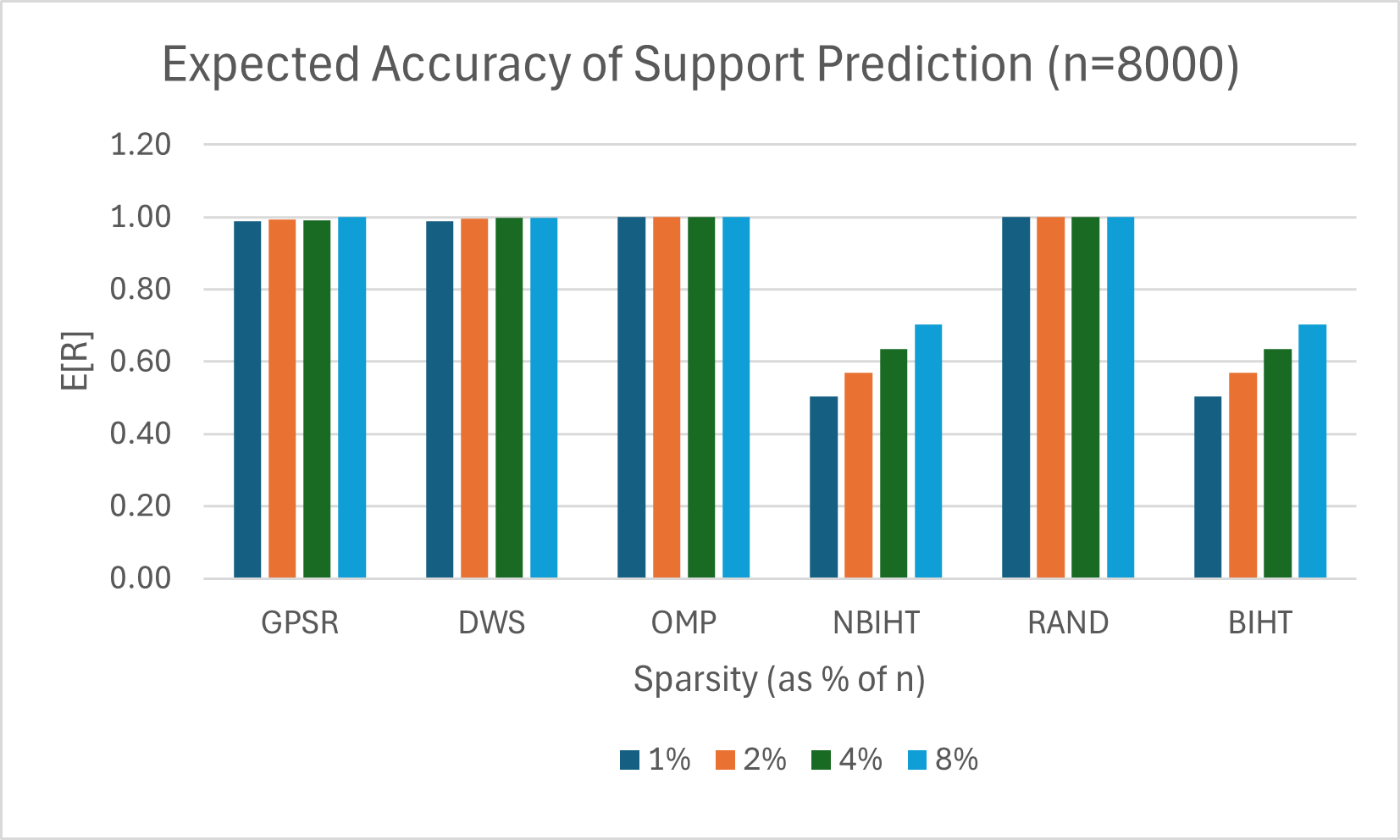}
        \subcaption{Performance for $n=8000$.}\label{fig:sp3}
    \end{minipage}
    
    \caption{
       Expected accuracy of support prediction between algorithms.
    }
    \label{fig:mainfig1}
\end{figure}

As shown in Fig.~\ref{fig:mainfig1}, GPSR, DWS, OMP, and RAND achieve high accuracy for the conventional compressed sensing problem. Among these, RAND and OMP perform best.  Note that OMP is given the sparsity \(s\) as an input argument, so OMP can ensure that \(|\supp(x)|\le s\). 

BIHT and NBIHT have the same level of accuracy for the harder 1-bit compressed sensing problem. As expected, since only the signs of the coordinates of $b$ are used by BIHT and NBIHT, its accuracy is much lower than the other methods that use the values of the coordinates of $b$.

\subsection{Speedup}

We define the speedup as the ratio of a baseline solver's running time to our running time, i.e., \(\text{speedup} = T_{\text{baseline}}/T_{\text{ours}}\). Thus, values greater than one indicate that our solver is faster, while values less than one indicate that it is slower.

Our method is between \(2.8\times\) and \(59.17\times\) faster than GPSR and between \(1.95\times\) and \(23.92\times\) faster than DWS. The advantage over the working-set methods is more pronounced when the solution is sparse. It is also between \(2.65\times\) and \(308.25\times\) faster than OMP, whose running time increases dramatically as \(s\) grows.

In an indirect comparison with one-bit compressed sensing solvers, our method is at least \(169\times\) faster than BIHT. Its speed is also comparable to NBIHT, with an average speedup of \(5.48\times\) overall; however, in the worst case the speedup drops to \(0.63\times\) for some larger \((n,s)\) settings.

\begin{figure}[htbp]
    \centering
    \begin{minipage}[b]{0.5\textwidth}
        \centering
        \includegraphics[width=0.9\linewidth]{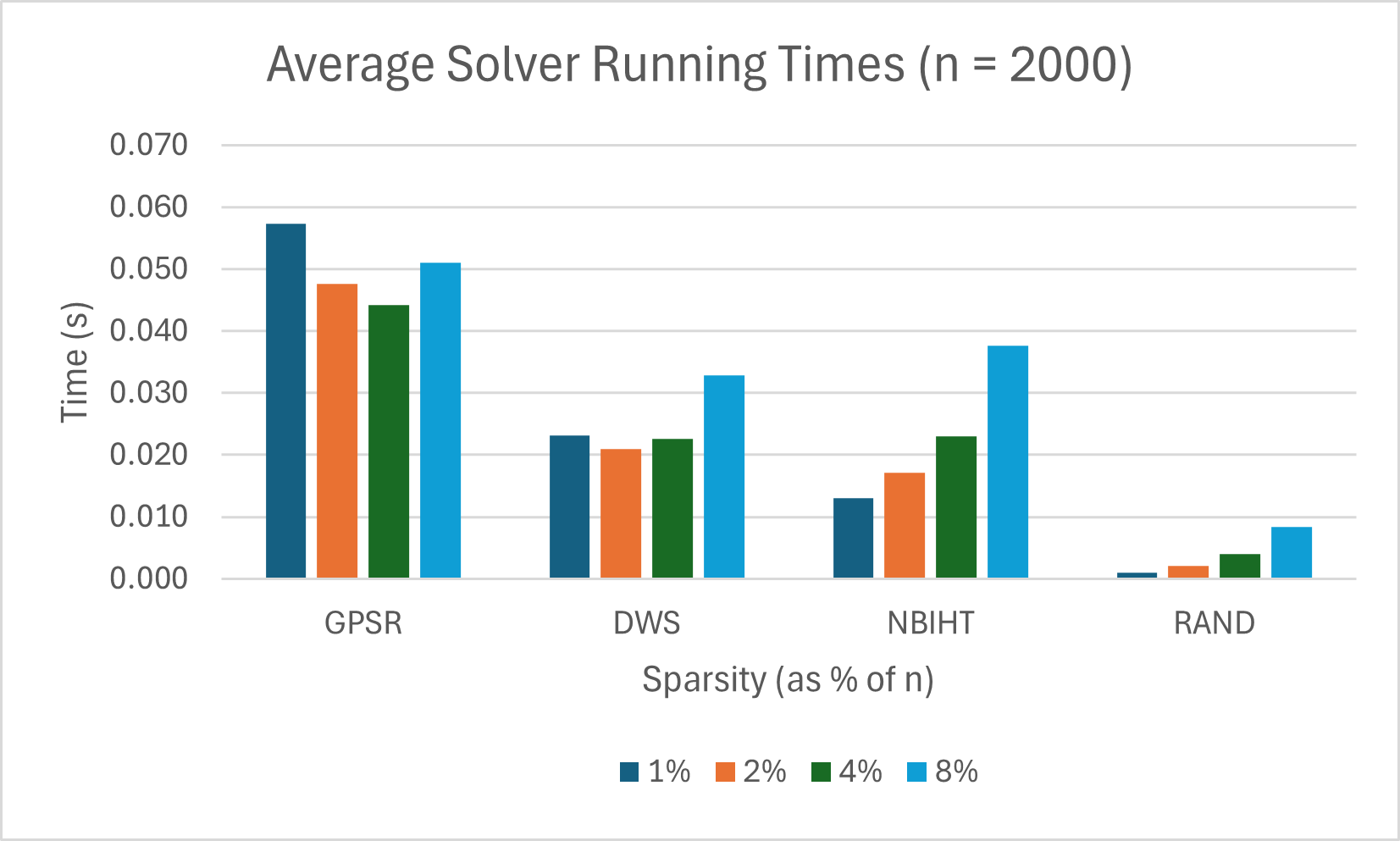}
        \subcaption{Performance for $n=2000$.}\label{fig:AM1}
    \end{minipage}
    \hfill
    \begin{minipage}[b]{0.5\textwidth}
        \centering
        \includegraphics[width=0.9\linewidth]{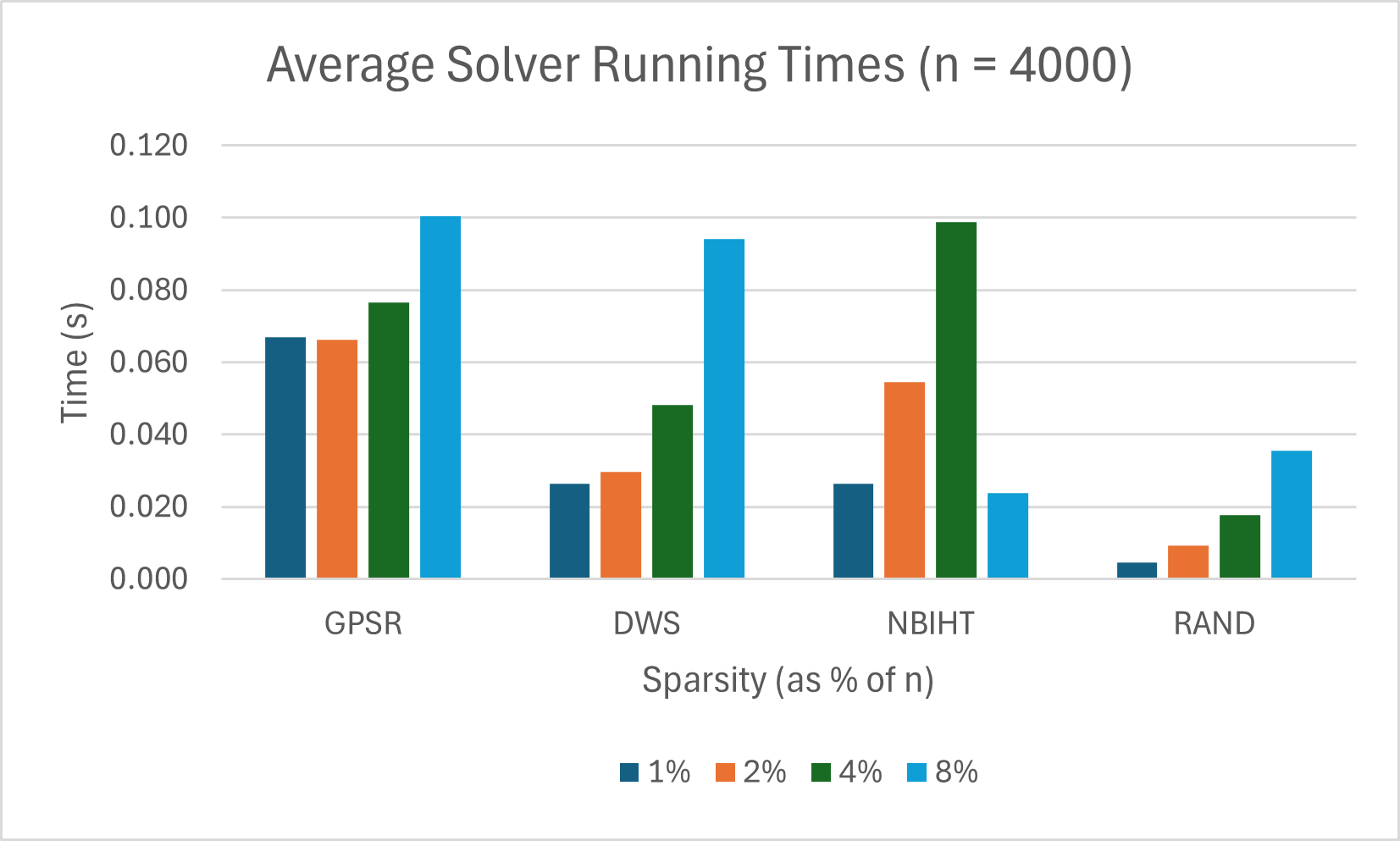}
        \subcaption{Performance for $n=4000$.}\label{fig:AM2}
    \end{minipage}
    \hfill
   \begin{minipage}[b]{0.5\textwidth}
        \centering
        \includegraphics[width=0.9\linewidth]{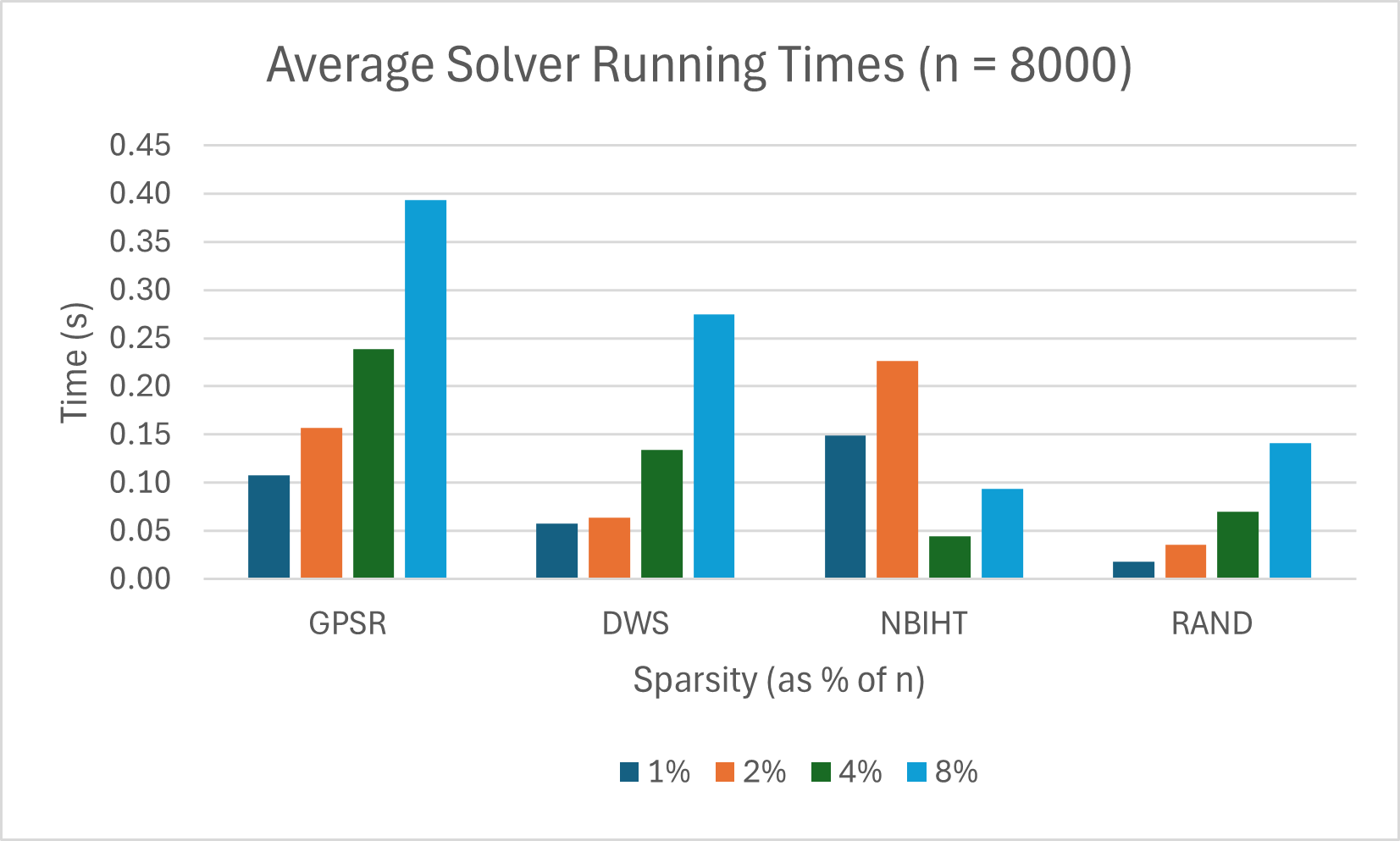}
        \subcaption{Performance for $n=8000$.}\label{fig:AM3}
    \end{minipage}
    
    \caption{
      Performance comparison between GPSR, DWS, NBIHT, and our method; the vertical axis shows runtime in seconds(s).
    }
    \label{fig:mainfig2}
\end{figure}

We summarize the results in Fig.~\ref{fig:mainfig2}, omitting OMP and BIHT because including them would distort the scale.

\newpage

\bibliographystyle{IEEEtran}
\bibliography{mybibliography}

\onecolumn
\appendix

\subsection{Proof of Lemma~\ref{lem:tech1}}

\noindent {\bf Lemma~\ref{lem:tech1}}~~
\emph{We write $v_i^{(r)}$ as $v_i$ for convenience.}
\begin{enumerate}[{(i)}]

\item $\forall\,\, i \in [n]$, $\ex[v_i] = z_i$.

\item $\forall\,\, i,j \in [n]$, if $i \not= j$, then $\cov(v_i,v_j) = 0$.

\item $\forall\,\, i \in [n]$, $\var[v_i] = \frac{1}{k}z_i^2 + \frac{1}{k}\norm{z}^2 + \sigma_w^2$.

\end{enumerate}

\vspace{4pt}

\begin{proof}
We omit the superscript ``(r)'' for convenience.
We have $v = A^\top b = A^\top A z + A^\top w$.  For $p \in [k]$, the $p$-th coordinate of $Az$ is $\sum_{q=1}^n a_{pq}z_q$.  Therefore, the $i$-th coordinate of $A^\top A z$ is $\sum_{p=1}^k a_{pi} \cdot \left(\sum_{q=1}^n a_{pq}z_q\right)$.  Similarly, the $i$-th coordinate of $A^\top w$ is $\sum_{p=1}^k a_{pi}w_p$.  Consequently,
\[
\forall\,\, i \in [n], \quad v_i = \sum_{p=1}^k\sum_{q=1}^n a_{pi}a_{pq}z_q + \sum_{p=1}^k a_{pi}w_p.
\]

Consider (i).
\begin{align*}
    \ex[v_i]  &= \sum_{p=1}^k \sum_{q=1}^n \ex[a_{pi}a_{pq}]z_q+ \sum_{p=1}^k  \ex[a_{pi}] \ex[w_p] \\
              &= \sum_{p=1}^k \sum_{q \neq i} \ex[a_{pi}]\ex[ a_{pq}]z_q + \sum_{p=1}^k \ex[a_{pi}^2]z_i  + \sum_{p=1}^k \ex[a_{pi}] \ex[w_p] \\
              &= \sum_{p=1}^k \var[a_{pi}] \cdot z_i  \quad\quad\quad (\because \ex[a_{pi}] = 0 \,\, \text{and} \,\, \var[a_{pi}] = \ex[a_{pi}^2]) \\
             &= z_i. \hspace{1.9in} (\because \var[a_{pi}] = 1/k) \\
\end{align*}

Consider (ii).
\begin{align*}
&~~~~\cov[v_i, v_j]  \\
&= \ex[v_iv_j]-\ex[v_i]\ex[v_j] \\
                &= \ex \left [\left(\sum_{p=1}^k \sum_{q=1}^n a_{pi}a_{pq}z_q + \sum_{p=1}^k a_{pi}w_p \right) \cdot \left(\sum_{s=1}^k \sum_{t=1}^n a_{sj}a_{st}z_t + \sum_{s=1}^k a_{sj}w_s \right)\right] -  z_i  z_j.
\end{align*}
The expected value of any term that involves either $w_p$ or $w_s$ is zero because $\ex[w_p] = \ex[w_s] = 0$.  The terms that involve both $w_p$ and $w_s$ are $a_{pi}a_{sj}w_pw_s$.  Since $i \not= j$, the variables $a_{pi}$ and $a_{sj}$ are independent no matter whether $p=s$ or not.  Therefore, $\ex[a_{pi}a_{sj}w_pw_s] = \ex[a_{pi}]\ex[a_{sj}]\ex[w_pw_s] = 0$ as $\ex[a_{pi}] = \ex[a_{sj}] = 0$.  We conclude that
\[
\cov[v_i,v_j] = \sum_{p=1}^k\sum_{s=1}^k\sum_{q=1}^n\sum_{t=1}^n  \ex\left[a_{pi}a_{pq}a_{sj}a_{st}\right]z_qz_t-  z_i  z_j.
\]
Suppose that $p \not= s$.  If $q \not= i$ or $t \not= j$, then $\ex[a_{pi}a_{pq}a_{sj}a_{st}] = \ex[a_{pi}]\ex[a_{pq}]\ex[a_{sj}a_{st}] = 0$ in the former case, and $\ex[a_{pi}a_{pq}a_{sj}a_{st}] = \ex[a_{pi}a_{pq}]\ex[a_{sj}][a_{st}] = 0$ in the latter case.  So only the case of $q=i$ and $t=j$ matters, yielding the term $\ex[a_{pi}^2a_{sj}^2]z_iz_j$. Similarly, if $p = s$, one can check that only the case of $q=i$ and $t=j$ matters.  Hence,
\[
\cov[v_i,v_j] = \sum_{p=1}^k \sum_{s=1}^k \ex[a_{pi}^2 a_{sj}^2] z_i z_j -  z_i  z_j.
\]
We have $\ex[a_{pi}^2] = \var[a_{pi}] + \ex[a_{pi}]^2 = 1/k$.  Since $i \not= j$, the variables $a_{pi}$ and $a_{sj}$ are independent, which implies that $\ex[a_{pi}^2a_{sj}^2] = \ex[a_{pi}^2]\ex[a_{sj}^2] = 1/k^2$.  It follows that
\[
\cov[v_i,v_j] = k^2 \cdot (z_iz_j/k^2) - z_iz_j = 0.
\]

Consider (iii).
\begin{align}
    \var[v_i] &= \ex\left[\left(\sum_{p=1}^k \sum_{q=1}^n a_{pi}a_{pq}z_q + \sum_{p=1}^k a_{pi}w_p\right)^2\right ]-\ex[v_i]^2 \nonumber \\
              &= \ex\left [\left(\sum_{p=1}^k \sum_{q=1}^n a_{pi}a_{pq}z_q\right)^2 \right ]+ 2\cdot \ex \left [\left(\sum_{p=1}^k \sum_{q=1}^n a_{pi}a_{pq}z_q\right)\left(\sum_{p=1}^k a_{pi}w_p\right) \right ] \nonumber \\
              &~~~~+ \ex \left [\left(\sum_{p=1}^k a_{pi}w_p\right)^2 \right ] - z_i^2.   \label{eq:1}
\end{align}
The first term in \eqref{eq:1} is equal to
\[
\sum_{p=1}^k\sum_{s=1}^k\sum_{q=1}^n\sum_{t=1}^n \ex[a_{pi}a_{pq}a_{si}a_{st}] \cdot z_qz_t.
\]
Suppose that $p \not= s$.  If $q \not= i$ or $t \not= i$, then $\ex[a_{pi}a_{pq}a_{si}a_{st}] = \ex[a_{pi}]\ex[a_{pq}]\ex[a_{si}a_{st}]$ or $\ex[a_{pi}a_{pq}]\ex[a_{si}]\ex[a_{st}]$, which is zero in either case.  Suppose that $p = s$.  If $q \not= t$, then $\ex[a_{pi}a_{pq}a_{si}a_{st}] = \ex[a_{pi}^2a_{pq}]\ex[a_{pt}]$ or $\ex[a_{pi}^2a_{pt}]\ex[a_{pq}]$, which is zero in either case.  Therefore,
\begin{align*}
\ex\left [\left(\sum_{p=1}^k \sum_{q=1}^n a_{pi}a_{pq}z_q\right)^2 \right]
&= \sum_{p=1}^k \sum_{q=1}^n \ex[a_{pi}^2a_{pq}^2] \cdot z_q^2 + \sum_{p=1}^k\sum_{s \not= p}\ex[a_{pi}^2a_{si}^2] \cdot z_i^2 \\
&= \sum_{p=1}^k\ex[a_{pi}^4] \cdot z_i^2 + \sum_{p=1}^k\sum_{q\not= i}\ex[a_{pi}^2]\ex[a_{pq}^2]\cdot z_q^2 \\
&~~~~+ \sum_{p=1}^k\sum_{s\not=p} \ex[a_{pi}^2]\ex[a_{si}^2] \cdot z_i^2.
\end{align*}
It is well known that $\ex[a_{pi}^4] = 3\cdot\var[a_{pi}]^2 = 3/k^2$~\cite{S2012}.  Recall that $\ex[a_{pi}^2] = \var[a_{pi}] + \ex[a_{pi}]^2 = 1/k$.  Hence,
\begin{align}
\ex\left [\left(\sum_{p=1}^k \sum_{q=1}^n a_{pi}a_{pq}z_q\right)^2 \right]
&= 3z_i^2/k + \sum_{q\not=i} z_q^2/k + (k-1)z_i^2/k \nonumber \\
&= \norm{z}^2/k + (k+1)z_i^2/k.  \label{eq:2}
\end{align}
Every subterm in the second term in \eqref{eq:1} involves a single occurrence of $w_p$.  Since $\ex[w_p] = 0$, we have
\begin{equation}
2\cdot \ex \left [\left(\sum_{p=1}^k \sum_{q=1}^n a_{pi}a_{pq}z_q\right)\left(\sum_{p=1}^k a_{pi}w_p\right) \right ] = 0.  \label{eq:3}
\end{equation}
The third term in~\eqref{eq:1} is equal to
\[
\sum_{p=1}^k\sum_{s=1}^k \ex[a_{pi}a_{si}] \ex[w_pw_s].
\]
If $p \not= s$, then $\ex[a_{pi}a_{si}] = \ex[a_{pi}]\ex[a_{si}] = 0$.  If $p=s$, then $\ex[a_{pi}^2] = 1/k$ and $\ex[w_p^2] = \sigma_w^2$, which makes 
\begin{equation}
\sum_{p=1}^k\sum_{s=1}^k \ex[a_{pi}a_{si}] \ex[w_pw_s] = \sigma_w^2.   \label{eq:4}
\end{equation}  
Putting \eqref{eq:1}--\eqref{eq:4} together gives $\var[v_i] = z_i^2/k + \norm{z}^2/k + \sigma_w^2$.
\end{proof}

\subsection{Proof of Lemma~\ref{lem:tech2}}

\vspace*{6pt}

\noindent {\bf Lemma~\ref{lem:tech2}}~~We write $b^{(r)}$ as $b$ for convenience.
\begin{enumerate}[(i)]

\item $\ex\bigl[\norm{b}^2\bigr] = \norm{z}^2 + k\sigma_w^2$.

\item $\var\bigl[\norm{b}^2\bigr] = \frac{2}{k}\left(\norm{z}^2 + k\sigma_w^2\right)^2$.

\end{enumerate}
\emph{Let $\sigma^2 = \mathrm{median}\{\norm{b^{(r)}}^2 : r \in [r_0]\}$.  It holds with probability more than $1-2e^{-r_0/72}$ that $\sigma^2$ lies strictly between $(1 -\sqrt{6/k}\bigr)\cdot\left(\|z\|^2 + k\sigma_w^2\right)$ and $\bigl(1 + \sqrt{6/k}\bigr) \cdot \left(\|z\|^2 + k\sigma_w^2\right)$.}

\vspace{6pt}

\begin{proof}
For any $q \not= t$, $a_{iq}z_q$ and $a_{it}z_t$ are independent random variables. Therefore, the set of random variables $\bigl\{a_{pq}z_q : q \in [n]\bigr\}$ are mutually independent.  As $a_{pq} \sim \mathcal{N}(0,1/k)$, we have $a_{pq}z_q \sim \mathcal{N}(0,z_q^2/k)$. Observe that $b_i = \sum_{q=1}^n a_{iq}z_q + w_i$.

Consider (i).
\begin{align*}
\ex\left[b_i^2\right] &= \ex\left[\left(\sum_{q=1}^n a_{iq}z_q + w_i\right)^2\right]\\
&= \ex\left[\sum_{q=1}^n\sum_{t=1}^n a_{iq}z_q a_{it}z_t\right] + \ex[w_i^2] +\ex\left[2w_i\cdot \sum_{q=1}^n a_{iq}z_q\right].
\end{align*}
In the first term, if $q \not= t$, then $\ex[a_{iq}z_qa_{it}z_t] = \ex[a_{iq}z_q] \cdot \ex[a_{it}z_t] = 0$ as $\ex[a_{iq}z_q]=\ex[a_{it}z_t]=0$.  The third term is zero as $\ex[w_i] = 0$.  Therefore, 
\[
\ex\left[b_i^2\right] = \sum_{q=1}^n \ex[a_{iq}^2z_q^2] + \ex[w_i^2] 
= \sum_{q=1}^n \var[a_{iq}z_q] + \var[w_i] \\
= \sum_{q=1}^n z_q^2/k + \sigma_w^2 = \norm{z}^2/k + \sigma_w^2.
\]
Consequently, $\ex\left[\norm{b}^2\right] = \sum_{i=1}^k \ex\left[b_i^2\right] = \norm{z}^2 + k\sigma_w^2$.  This completes the proof of (i).

Consider(ii).  Since $b_i = \sum_{q=1}^n a_{iq}z_q + w_i$, we have
\[
\ex[b_i] = \sum_{q=1}^n \ex[a_{iq}z_q] + \ex[w_i] = 0,
\]
\begin{align*}
\var\bigl[b_i\bigr] 
&= \var\left[\sum_{q=1}^n a_{iq}z_q\right] +\var[w_i] \,\, = \,\, \sum_{q=1}^n \var[a_{iq}z_q] + \var[w_i] \,\, = \,\, \sum_{q=1}^n z_q^2/k + \sigma_w^2 \\
&=\norm{z}^2/k + \sigma_w^2.
\end{align*}
The sum of Gaussian variables is a Gaussian variable, which makes $b_i$ a Gaussian variable with zero mean and variance $\norm{z}^2/k + \sigma_w^2$.  In this case, it is well known~\cite{S2012} that
\[
\ex\left[b_i^4\right] = 3\cdot\var[b_i]^2 = 3\left(\norm{z}^2/k + \sigma_w^2\right)^2.
\]
It follows that
\[
\var\left[b_i^2\right] = \ex\left[b_i^4\right] - \ex\left[b_i^2\right]^2
= 3\left(\norm{z}^2/k + \sigma_w^2\right)^2 - \left(\norm{z}^2/k + \sigma_w^2\right)^2 = 2\left(\norm{z}^2/k + \sigma_w^2\right)^2.
\]
For any $i, j \in [k]$, if $i \not= j$, then $b_i = \sum_{q=1}^n a_{iq}z_q + w_i$ and $b_j = \sum_{q=1}^n a_{jq}z_q + w_j$ are clearly independent random variables.  It follows that $b_i^2$ and $b_j^2$ are also independent.  Consequently,
\[
\var\left[\norm{b}^2\right] = \sum_{i=1}^k \var\left[b_i^2\right] = 2k\left(\norm{z}^2/k + \sigma_w^2\right)^2 = \frac{2}{k}\left(\norm{z}^2 + k\sigma_w^2\right)^2.
\]
This completes the proof of (ii).

Define a random binary variable $\beta^{(r)}$ such that $\beta^{(r)} = 1$ if and only if $\norm{b^{(r)}}^2$ deviates less than $\sqrt{6/k} \cdot \left(\norm{z}^2 + k\sigma_w^2\right)$ from $\norm{z}^2 + k\sigma_w^2$.  By Chebyshev's inequality, it happens with probability at most $1/3$ that $\norm{b^{(r)}}^2$ deviates from $\norm{z}^2 + k\sigma_w^2$ by at least $\sqrt{6/k}\left(\norm{z}^2 + k\sigma_w^2\right)$.  As a result, $\Pr\left[\beta^{(r)}=1\right] \geq 2/3$, which implies that $\ex\left[\beta^{(r)}\right] \geq 2/3$.

Let $y=\sum_{r=1}^{r_0} \beta^{(r)}$.  By Chernoff bound, it holds with probability at most $2e^{-\ex[y]/48}$ that $y$ deviates from $\ex[y]$ by at least $\ex[y]/4$.   Since $\ex\left[\beta^{(r)}\right] \geq 2/3$, we have $\ex[y] \geq 2r_0/3$.  It follows that $y \leq r_0/2 \Rightarrow y \leq 3\ex[y]/4$, which implies that $\Pr[y\leq r_0/2] \leq 2e^{-\ex[y]/48} \leq 2e^{-r_0/72}$. When $y > r_0/2$, it means that $\norm{b^{(r)}}^2$ deviates from its mean $\norm{z}^2 + k\sigma_w^2$ by less than $\sqrt{6/k} \left(\norm{z}^2 + k\sigma_w^2\right)$ in $\lceil r_0/2 \rceil$ trials or more.  Hence, it holds with probability more than $1-2e^{-r_0/72}$ that $\sigma^2$ lies strictly between $(1 -\sqrt{6/k}\bigr)\cdot\left(\|z\|^2 + k\sigma_w^2\right)$ and $\bigl(1 + \sqrt{6/k}\bigr) \cdot \left(\|z\|^2 + k\sigma_w^2\right)$.
\end{proof}

\subsection{Proof of Lemma~\ref{lem:class1}}

\vspace*{6pt}

\noindent {\bf Lemma~\ref{lem:class1}}~~\emph{Assume that $k \geq 54$, $\sigma > (1-\sqrt{6/k})^{1/2}\cdot \sqrt{\norm{z}^2 + k\sigma_w^2}$, and $\min_{i \in \supp(z)} |z_i| \geq 6\sigma/\sqrt{k}$.  It holds with probability at least $1-2ne^{-r_0/540}$ that for $i \in [n]$, if $i \not\in \supp(z)$, then $\hat{z}_i < 2\sigma/\sqrt{k}$; otherwise, $\hat{z}_i > 2\sigma/\sqrt{k}$.}

\vspace{6pt}

\begin{proof}
We first deal with the case that $i \not\in \supp(z)$.  Let $\sigma_i = (\var[v_i^{(r)}])^{1/2}$.  By Lemma~\ref{lem:tech1}(iii), $\frac{3}{2}\sigma_i = \frac{3}{2\sqrt{k}}\sqrt{z_i^2+ \norm{z}^2+ k\sigma_w^2} \leq \frac{3}{2\sqrt{k}}z_i + \frac{3}{2\sqrt{k}}\sqrt{\norm{z}^2 + k\sigma_w^2}$.

By Lemma~\ref{lem:tech1} and Chebyshev's inequality, $v_i^{(r)}$ deviates from its mean $z_i$ by at least $\frac{3}{2}\sigma_i$ with probability at most $4/9$.  Therefore, it holds with probability at least $5/9$ that $v_i^{(r)}$ deviates from 
its mean $z_i$ by less than $\frac{3}{2}\sigma_i \leq \frac{3}{2\sqrt{k}}z_i + \frac{3}{2\sqrt{k}}\sqrt{\norm{z}^2 + k\sigma_w^2}$.  For $i \not\in \supp(z)$, $z_i = 0$, which means that $\bigl|v_i^{(r)}\bigr| < \frac{3}{2\sqrt{k}}\sqrt{\norm{z}^2 + k\sigma_w^2}$ with probability at least 5/9. 

Define a random binary variable $\beta^{(r)}$ such that $\beta^{(r)} = 1$ if and only if $\bigl|v_i^{(r)}\bigr| < \frac{3}{2\sqrt{k}}\sqrt{\norm{z}^2 + k\sigma_w^2}$.  Therefore, $\ex\left[\beta^{(r)}\right] \geq 5/9$. Let $y =\sum_{r=1}^{r_0} \beta^{(r)}$.   By Chernoff bound, $y$ deviates from $\ex[y]$ by at least $\ex[y]/10$ with probability at most $2e^{-\ex[y]/300}$.  Since $\ex\left[\beta^{(r)}\right] \geq 5/9$, we have $\ex[y] \geq 5r_0/9$.  It follows that $y \leq r_0/2 \Rightarrow y \leq 9\ex[y]/10$ which happens with probability at most $2e^{-\ex[y]/300} \leq 2e^{-r_0/540}$.  When $y \ge r_0/2$, it means that $\bigl|v_i^{(r)}\bigr|  < \frac{3}{2\sqrt{k}}\sqrt{\norm{z}^2 + k\sigma_w^2}$ in $\lceil r_0/2 \rceil$ trials or more.

By assumption, $\sigma > (1-\sqrt{6/k})^{1/2} \cdot \sqrt{\norm{z}^2+k\sigma_w^2)}$, which is at least $\sqrt{2/3} \cdot \sqrt{\norm{z}^2 + k\sigma_w^2}$ as $k \geq 54$.  Hence, $\bigl|v_i^{(r)}\bigr| < (3/2)^{1.5}\sigma/\sqrt{k} < 2\sigma/\sqrt{k}$ in $\lceil r_0/2 \rceil$ trials or more with probability at least $1 - 2e^{-r_0/540}$.   By the union bound, it holds with probability at least $1 - 2(n-s)e^{-r_0/540}$ that $|\hat{z}_i| < 2\sigma/\sqrt{k}$ for all $i \not\in \supp(z)$.

Next, we deal with case that $i \in \supp(z)$.  We have shown previously that $v_i^{(r)}$ deviates from its mean $z_i$ by less than $\frac{3}{2\sqrt{k}}z_i + \frac{3}{2\sqrt{k}}\sqrt{\norm{z}^2 + k\sigma_w^2}$ with probability at least 5/9, and that $\sigma > \sqrt{2/3} \cdot \sqrt{\norm{z}^2 + k\sigma_w^2}$.  Therefore, $v_i^{(r)}$ deviates from $z_i$ by less than $\frac{3}{2\sqrt{k}}z_i + \frac{(3/2)^{1.5}}{\sqrt{k}}\sigma < \frac{3}{2\sqrt{k}}z_i + \frac{2}{\sqrt{k}}\sigma$ with probability at least $5/9$.  By the assumptions of $k \geq 54$ and $\min_{i \in \supp(z)} |z_i| \geq 6\sigma/\sqrt{k}$, it holds with probability at least 5/9 that  $\bigl|v_i^{(r)}\bigr| > \left(1-\frac{3}{2\sqrt{k}}\right)|z_i| -2\sigma/\sqrt{k} > 4\sigma/\sqrt{k} - 2\sigma/\sqrt{k} = 2\sigma/\sqrt{k}$.

Define a random binary variable $\beta^{(r)}$ such that $\beta^{(r)} = 1$ if and only if $\bigl|v_i^{(r)}\bigr| > 2\sigma /\sqrt{k}$. Therefore, $\ex[\beta^{(r)}] \ge 5/9$. Let $y=\sum_{r=1}^{r_0} \beta^{(r)}$. Then, $\ex[y] \geq 5r_0/9$. By Chernoff bound, $y$ deviates from $\ex[y]$ by at least $\ex[y]/10$ with probability at most $2e^{-\ex[y]/300} \leq 2e^{-r_0/540}$.  Note that $y \leq r_0/2 \Rightarrow y \leq 9\ex[y]/10$.  So $\Pr[y \leq r_0/2] \leq 2e^{-r_0/540}$, which means that $\bigl|v_i^{(r)}\bigr| > 2\sigma /\sqrt{k}$ in $\lceil r_0/2 \rceil$ trials or more with probability at least $1-2e^{-r_0/540}$. By the union bound, it holds with probability at least $1-2se^{-r_0/540}$ that $|\hat{z}_i| > 2\sigma/\sqrt{k}$ for all $i \in \supp(z)$.
\end{proof}

\end{document}